\documentclass[aps,prl,twocolumn,superscriptaddress]{revtex4-2}
\usepackage{graphicx}
\usepackage{bm}
\usepackage{xcolor}
\usepackage{amsmath}
\usepackage{amssymb}
\usepackage{hyperref}

\begin{document}

\title{Possible Dirac quantum spin liquid in the kagome quantum antiferromagnet YCu$_3$(OH)$_6$Br$_2$[Br$_{x}$(OH)$_{1-x}$]}

\author{Zhenyuan Zeng}
\affiliation{Beijing National Laboratory for Condensed Matter Physics, Institute of Physics, Chinese Academy of Sciences, Beijing 100190, China}
\affiliation{School of Physical Sciences, University of Chinese Academy of Sciences, Beijing 100190, China}
\author{Xiaoyan Ma}
\affiliation{Beijing National Laboratory for Condensed Matter Physics, Institute of Physics, Chinese Academy of Sciences, Beijing 100190, China}
\affiliation{School of Physical Sciences, University of Chinese Academy of Sciences, Beijing 100190, China}
\author{Si Wu}
\affiliation{Institute of Applied Physics and Materials Engineering, University of Macau, Avenida da Universidade, Taipa, Macao SAR 999078, China}
\author{Hai-Feng Li}
\email{haifengli@um.edu.mo}
\affiliation{Institute of Applied Physics and Materials Engineering, University of Macau, Avenida da Universidade, Taipa, Macao SAR 999078, China}
\author{Zhen Tao}
\affiliation{Center for Advanced Quantum Studies and Department of Physics, Beijing Normal University, Beijing 100875, China}
\author{Xingye Lu}
\affiliation{Center for Advanced Quantum Studies and Department of Physics, Beijing Normal University, Beijing 100875, China}
\author{Xiao-hui Chen}
\affiliation{Fujian Provincial Key Laboratory of Advanced Materials(Xiamen University), Department of Materials Science and Engineering, College of Materials, Xiamen University, Xiamen 361005, China}
\author{Jin-Xiao Mi}
\affiliation{Fujian Provincial Key Laboratory of Advanced Materials(Xiamen University), Department of Materials Science and Engineering, College of Materials, Xiamen University, Xiamen 361005, China}
\author{Shi-Jie Song}
\affiliation{Department of Physics, Zhejiang University, Hangzhou 310027, China}
\author{Guang-Han Cao}
\affiliation{Department of Physics, Zhejiang University, Hangzhou 310027, China}
\affiliation{Zhejiang Province Key Laboratory of Quantum Technology and Devices, Interdisciplinary Center for Quantum Information, and State Key
Lab of Silicon Materials, Zhejiang University, Hangzhou 310027, China}
\affiliation{Collaborative Innovation Centre of Advanced Microstructures, Nanjing University, Nanjing 210093, China}
\author{Guangwei Che}
\affiliation{Center for High Pressure Science and Technology Advanced Research, 10 Xibeiwang East Road, Haidian, Beijing 100094, China}
\author{Kuo Li}
\affiliation{Center for High Pressure Science and Technology Advanced Research, 10 Xibeiwang East Road, Haidian, Beijing 100094, China}
\author{Gang Li}
\affiliation{Beijing National Laboratory for Condensed Matter Physics, Institute of Physics, Chinese Academy of Sciences, Beijing 100190, China}
\affiliation{Songshan Lake Materials Laboratory, Dongguan, Guangdong 523808, China}
\author{Huiqian Luo}
\affiliation{Beijing National Laboratory for Condensed Matter Physics, Institute of Physics, Chinese Academy of Sciences, Beijing 100190, China}
\affiliation{Songshan Lake Materials Laboratory, Dongguan, Guangdong 523808, China}
\author{Zi Yang Meng}
\affiliation{Beijing National Laboratory for Condensed Matter Physics, Institute of Physics, Chinese Academy of Sciences, Beijing 100190, China}
\affiliation{Department of Physics and HKU-UCAS Joint Institute of Theoretical and Computational Physics, The University of Hong Kong, Pokfulam Road, Hong Kong, China}
\author{Shiliang Li}
\email{slli@iphy.ac.cn}
\affiliation{Beijing National Laboratory for Condensed Matter Physics, Institute of Physics, Chinese Academy of Sciences, Beijing 100190, China}
\affiliation{School of Physical Sciences, University of Chinese Academy of Sciences, Beijing 100190, China}
\affiliation{Songshan Lake Materials Laboratory, Dongguan, Guangdong 523808, China}
\begin{abstract}
We studied the magnetic properties of YCu$_3$(OH)$_6$Br$_2$[Br$_{1-x}$(OH)$_{x}$] ($x$ = 0.33), where Cu$^{2+}$ ions form two-dimensional kagome layers. There is no magnetic order down to 50 mK while the Curie-Weiss temperature is on the order of -100 K. At zero magnetic field, the low-temperature specific heat shows a $T^2$ dependence. Above 2 T, a linear temperature dependence term in specific heat emerges, and the value of $\gamma = C/T$ increases linearly with the field. Furthermore, the magnetic susceptibility tends to a constant value at $T = 0$. Our results suggest that the magnetic ground state of YCu$_3$(OH)$_6$Br$_2$[Br$_{1-x}$(OH)$_{x}$] is consistent with a Dirac quantum-spin-liquid state with a linearly dispersing spinon strongly coupled to an emergent gauge field, which has long been theoretically proposed as a candidate ground state in the two-dimensional kagome Heisenberg antiferromagnetic system. 
\end{abstract}

\maketitle

The two-dimensional (2D) kagome Heisenberg antiferromagnet (KHA) has stimulated great research interest and activities~\cite{SachdevS92,HastingsMB00,BalentsL02,BalentsL10,NormanMR16,SavaryL17,ZhouY17,BroholmC20}. This is because the Lieb-Schultz-Mattis-Oshikawa-Hastings theorems~\cite{Lieb1961,Oshikawa2000,Hastings2004} forbid KHA systems from having a trivially gapped ground state and therefore increase the possibilities of realizing exotic states such as quantum spin liquids (QSLs) and valence bond solids ~\cite{ChalkerJT92,RitcheyI93,YildirimT06,ZhouY17,Matan2010,Smaha2020}. If we consider only  the homogeneous nearest-neighbor exchange, the leading candidates for the ground states are gapped $Z_2$ QSL \cite{YanS11,DepenbrockS12,JiangHC12,FanY12,MeiJW17} and gapless U(1) Dirac QSL \cite{RanY07,HermeleM08,IqbalY11,IqbalY14,HeYC17,LiaoHJ17}. Introducing second-neighbor and third-neighbor exchanges or more generic interactions will result in even richer ground states \cite{HeYC14,GongSS15,GongSS16,WangYC18,GYSun2018,LeeCY18,PrelovsekP21}. Given these varieties, it is generally expected that a kagome magnetic material exhibiting no magnetic ordering at low temperatures is more likely to acquire a QSL ground state.

Experimentally, the best candidate so far for a kagome QSL is undoubtedly herbertsmithite [ZnCu$_3$(OH)$_6$Cl$_2$] \cite{ShoresMP05,HanTH12}, which consists of perfect kagome planes formed by Cu$^{2+}$ ions ($S$ = 1/2). The central debate for this material, if the ground state is indeed a QSL, is whether the spin system is gapped or gapless \cite{FuM15,HanTH16,KhuntiaP20}. While this is supposed to be easy to observe experimentally, the existence of a few percent of magnetic Cu$^{2+}$ ions on the nonmagnetic Zn$^{2+}$ sites makes the low-energy spectrum and low-temperature thermodynamical properties dominated by these impurity spins \cite{VriesMA08,FreedmanDE10,YYHuang2021}. Similar issues have also been found for many other KHAs \cite{KermarrecE11,LIY14,FengZL17,WeiYuan2017,FengZL18b,YingFu2021}. To avoid site disorder, YCu$_3$(OH)$_6$Cl$_3$ was synthesized with perfect Cu$^{2+}$ kagome planes but no site mixing because of very different ionic sizes of Y$^{3+}$ and Cu$^{2+}$ \cite{SunW16,PuphalP17}. However, it was later found that this compound may exhibit an antiferromagnetic (AFM) order at low temperatures \cite{ZorkoA19,ZorkoA19b,BarthelemyQ19}. Moreover, the samples are found to always contain clinoatacamite [Cu$_2$(OH)$_3$Cl] impurities, which may give rise to contradictory results \cite{SunW16,PuphalP17,ZorkoA19,ZorkoA19b}. Recently, a new compound, YCu$_3$(OH)$_6$Br$_2$[Br$_x$(OH)$_{1-x}$] ($x \approx$ 0.51), was reported which shows no magnetic ordering down to 2 K and seems to be a better QSL candidate \cite{ChenXH20}.

In this work, we synthesize high-quality single crystals of YCu$_3$(OH)$_6$Br$_2$[Br$_x$(OH)$_{1-x}$] ($x$ = 0.33). Few magnetic impurities are found, and their effects on the thermodynamical properties are negligible. The system shows no magnetic ordering down to 50 mK despite the large Curie-Weiss temperature ($\sim$ -79 K). The low-temperature specific heat $C$ is proportional to $T^2$ below 0.7 K, while the field-induced $T$-linear component of $C$ is proportional to the magnetic field. These results are consistent with the theoretical expectation of a Dirac QSL~\cite{RanY07}.

Single crystals of YCu$_3$(OH)$_6$Br$_2$[Br$_{x}$(OH)$_{1-x}$] (YCu3-H) were grown using the hydrothermal method reported previously \cite{ChenXH20}. The crystals are hexagonal plates with an in-plane diameter of 0.5 to 1 mm and a thickness of 0.1 to 0.3 mm. All the crystals were ultrasonically cleaned in water before measurements were taken to remove possible impurities attached to the surfaces of the crystals. The deuterated single crystals (YCu3-D) were synthesized using the same method with the corresponding deuterated starting materials and heavy water. All the results are for YCu3-H if not otherwise mentioned. The crystal structure and chemical formula were determined by single-crystal x-ray diffraction (SCXRD). Specific-heat and magnetic-susceptibility measurements were measured on physical property measurement systems (Quantum Design)  and magnetic property measurement systems (Quantum Design), respectively. To obtain a good enough signal, we typically used either $c$-axis-aligned or randomly oriented crystals in these measurements.

\begin{table}
  \centering
    \begin{tabular}{ccccccc}
        \hline
             Atom & $x$ & $y$ & $z$ & Occupancy & $U_{eq}$ (\AA$^2$)\\
        \hline
				Y11&0.0000&0.0000&0.5000&0.301(2)&0.0107(4)\\
				Y12&0.0000&0.0000&0.6222(3)&0.3493(12)&0.0107(4)\\
				Cu&0.5000 & 0.5000 & 0.5000 & 1 & 0.0127(3)\\
				Br1&0.666667&0.333333&0.85645(10)&1&0.0179(3)\\
				O1&0.1888(3) & 0.8112(3) & 0.6288(7) & 1 & 0.0298(8)\\
				Br2&0.0000 & 0.0000 & 0.0000 & 0.326(6) & 0.0207(8)\\
			  O2&0.0000 & 0.0000 & 0.0000 & 0.674(6) & 0.0207(8)\\

        \hline
    \end{tabular}
  \caption{Fractional atomic coordinates and equivalent isotropic displacement parameters of YCu$_3$(OH)$_6$[Br$_{0.33}$(OH)$_{0.67}$] with space group $P\bar{3}m1$ (No. 164): $a = b = 6.6784(2) $\AA, $c = 5.9901(3) $\AA, $\alpha = \beta = 90 ^{\circ}$, $\gamma = 120 ^{\circ}$.}
  \label{table1}
\end{table}

\begin{table}
  \centering
    \begin{tabular}{ccccccc}
        \hline
             Atom & x & y & z & Occupancy & $U_{eq}$ (\AA$^2$)\\
        \hline
				Y11&0.0000&0.0000&0.5000&0.314(2)&0.0076(4)\\
				Y12&0.0000&0.0000&0.6227(3)&0.3428(12)&0.0076(4)\\
				Cu&0.5000 & 0.5000 &0.5000 & 1 & 0.0099(3)\\
				Br1&0.666667&0.333333&0.85614(10)&1&0.048(3)\\
				O1&0.1890(7) & 0.8110(3) & 0.6298(7) & 1 & 0.0267(8)\\
				Br2&0.0000 & 0.0000 & 0.0000 & 0.317(8) & 0.0152(10)\\
			  O2&0.0000 & 0.0000 & 0.0000 & 0.683(8) & 0.0152(10)\\

        \hline
    \end{tabular}
  \caption{Fractional atomic coordinates and equivalent isotropic displacement parameters of YCu$_3$(OD)$_6$[Br$_{0.32}$(OH)$_{0.68}$] with the space group of $P\bar{3}m1$ (No. 164): $a = b =6.6779(3) $\AA, $c = 5.9874(4) $\AA, $\alpha = \beta = 90 ^{\circ}$, $\gamma = 120 ^{\circ}$. }
  \label{table2}
\end{table}

\begin{figure}[tbp]
\includegraphics[width=\columnwidth]{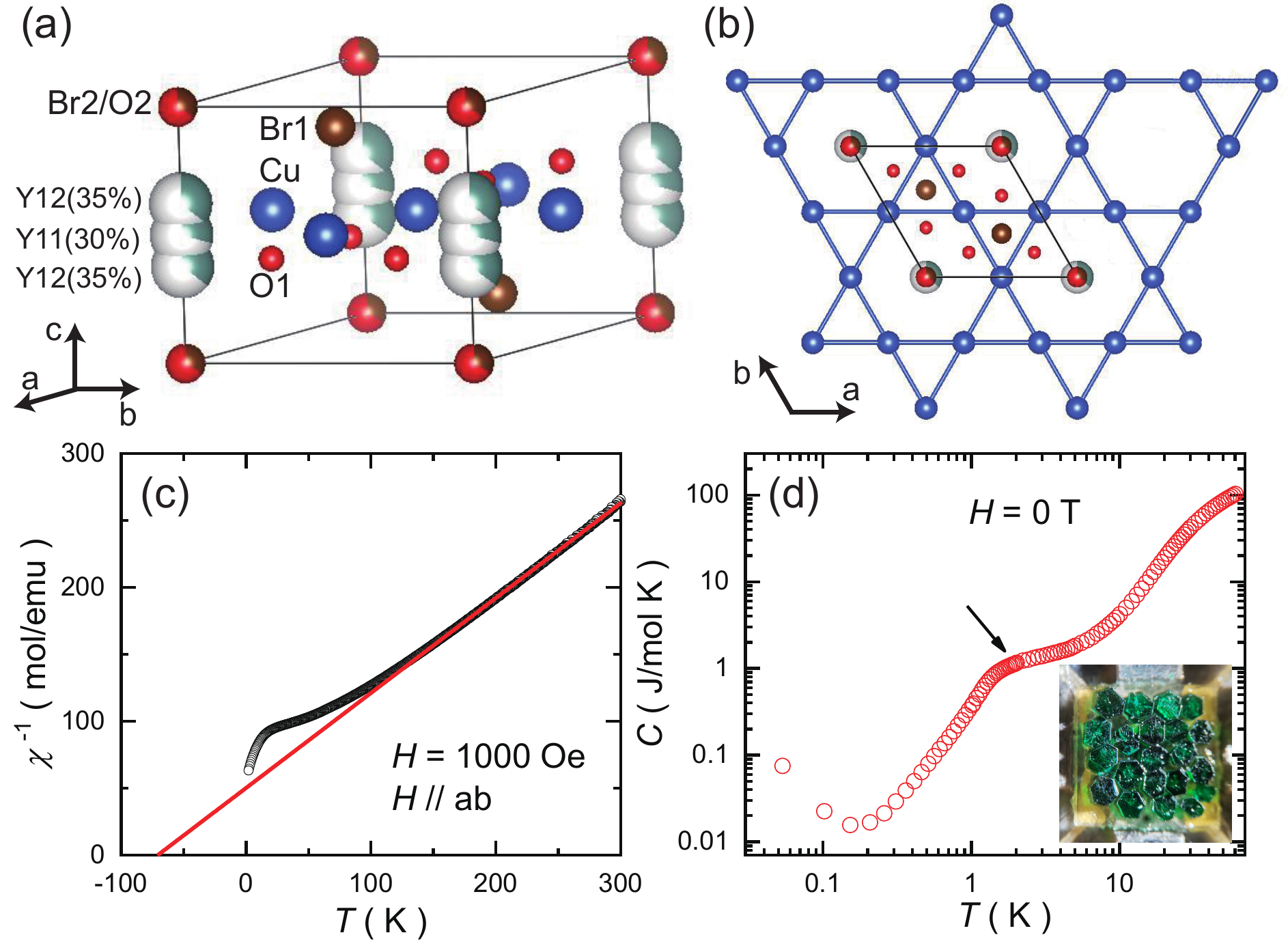}
 \caption{(a) The crystal structure of YCu$_3$(OH)$_6$[Br$_{0.33}$(OH)$_{0.67}$] where hydrogens are not shown. The solid lines represent the unit cell. (b) The kagome plane of Cu$^{2+}$ ions viewed along the $c$-axis. The other atoms are only shown in one unit cell. (c) The temperature dependence of the inverse of the magnetic susceptibility $\chi^{-1}$. The solid line is a linear fit for the high-temperature data. (d) The temperature dependence of the specific heat $C$ in the log-log scale. The arrow indicates the shoulder of the low-temperature specific heat. The inset shows a photo of several samples on the platform of the specific-heat puck.}
 \label{Structure}
\end{figure}

Figure \ref{Structure}(a) shows the crystal structure of YCu3-H ($x$ = 0.33), with detailed information given in Table \ref{table1}, which was obtained from SCXRD. As reported previously for the $x$ = 0.51 sample \cite{ChenXH20}, the Cu$^{2+}$ ions form 2D kagome layers [Fig. \ref{Structure}(b)]. Compared to the $x$ = 0.51 sample, the occupancy of Y11 increases from 10\% to 30\% while that of Br2 decreases from 51\% to 33\%. Moreover, the in-plane and $c$-axis lattice constants become slightly larger and smaller, respectively, and the Cu-O1-Cu angle for our sample (115.81$^{\circ}$) is larger than that for the $x$ = 0.51 sample (114.08$^{\circ}$). For YCu3-D ($x$ = 0.32), the results are almost the same, as shown in Table \ref{table2}. We note that the molecular formula for our sample may be roughly written as Y$_3$Cu$_9$(OH)$_{20}$Br$_{7}$, but no superstructure as reported in Refs.~\cite{PuphalP17,BarthelemyQ19,note1} has been found here. 

Figure \ref{Structure}(c) shows the inverse magnetic susceptibility $\chi^{-1} = H/M$ of YCu3-H as a function of the temperature for the field parallel to the $ab$ plane. A linear fit to the data above 150 K gives the Curie-Weiss temperature $\theta_{CW}$ and effective moment $\mu_{eff}$ as about -79 K and 1.94 $\mu_B$, respectively, which are similar to those for the $x$ = 0.51 sample \cite{ChenXH20}. It should be noted that these values will change when adding a temperature-independent background as a fitting parameter in the Curie-Weiss function, but $\theta_{CW}$ is always on the order of -100 K, suggesting large AFM interactions. 

\begin{figure}[tbp]
\includegraphics[width=\columnwidth]{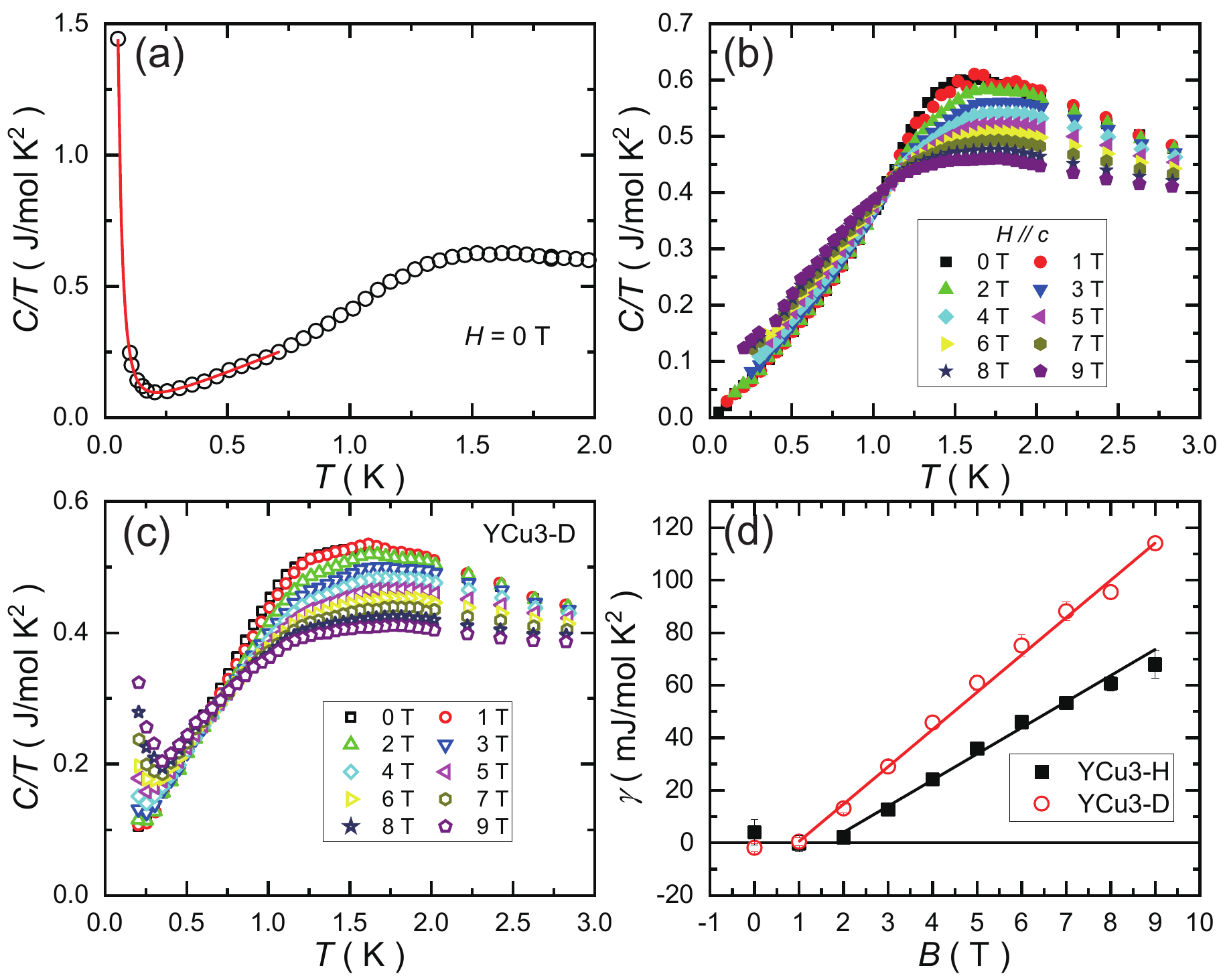}
 \caption{(a) The temperature dependence of $C/T$ below 2 K at 0 T. The solid line is the fitted result as described in the main text. (b) and (c) The temperature dependence of $C/T$ at different fields for YCu3-H and YCu3-D, respectively. (d) The field dependence of $\gamma$. The solid lines are fitted results with the linear function.}
 \label{SC}
\end{figure}

Figure \ref{Structure}(d) shows the temperature dependence of the specific heat, where no magnetic ordering is found down to 50 mK. A shoulder appears at about 2 K, which is typical for a KHA from numerical calculations~\cite{SchnackJ18}. One of the key questions now is whether the spin system is gapped or not. To see this more clearly, we plot $C/T$ below 2 K at zero field in Fig. \ref{SC}(a). The sharp upturn below about 0.2 K is from the nuclear Schottky anomaly and can be described by $AT^{-3}$, where $A$ is the fitting parameter. We find that no gap function can describe the data. Rather, the data below 0.7 K can be well fitted by $AT^{-3}+\alpha T^{n}$, with $n$ = 1 $\pm$ 0.02. Same measurements were done on several batches of samples and the value of $n$ is always close to 1. These results clearly show that the specific heat (after subtraction of the Schottky anomaly~\cite{note1}) is proportional to $T^2$ when $T$ goes to zero. For a Dirac QSL with a linearly dispersing spinon spectrum, such $T^2$ specific heat at zero field is expected~\cite{RanY07}. This is our first experimental evidence to show that the ground state of YCu3 may be a Dirac QSL.

Figure \ref{SC}(b) shows $C/T$ at different fields, where the contribution from the nuclear Schottky anomaly has been removed \cite{note1}. The hump at 1.5 K is gradually suppressed by the field while the specific heat below 1 K slightly increases with increasing field. This behavior rules out the existence of weakly correlated impurity spins that are typically found in herbertsmithite and many other kagome compounds \cite{VriesMA08,FreedmanDE10,KermarrecE11,LIY14,FengZL17,FengZL18b,YingFu2021,YYHuang2021}, whose contribution would be shifted to higher temperature under magnetic field \cite{VriesMA08}. Below about 0.7 K, $C/T$ shows a linear temperature dependence at all fields with little change in the slope. 

Figure \ref{SC}(c) further shows the results for the YCu3-D sample, where the nuclear Schottky anomaly becomes less significant. The linear temperature dependence of $C/T$ can thus be directly seen without the need to remove the Schottky contribution,  therefore avoiding the uncertainties at large fields. Interestingly, the slope clearly decreases with increasing field, which is different from that of YCu3-H. But for both the YCu3-H and YCu3-D samples, we can fit the $C/T$ data with a generic linear function, i.e., $C/T = \gamma + \alpha T$, and we show the obtained field dependence of $\gamma$ in Fig. \ref{SC}(d). We see that above 2 T, $\gamma$ for both samples increases linearly with the field, showing a linear temperature dependence of the specific heat at high fields. This is the second experimental evidence that the material exhibits consistent behavior of a Dirac QSL under magnetic field at low temperature, i.e., $k_B T \ll \mu_B B$~\cite{RanY07}.

\begin{figure}[tbp]
\includegraphics[width=\columnwidth]{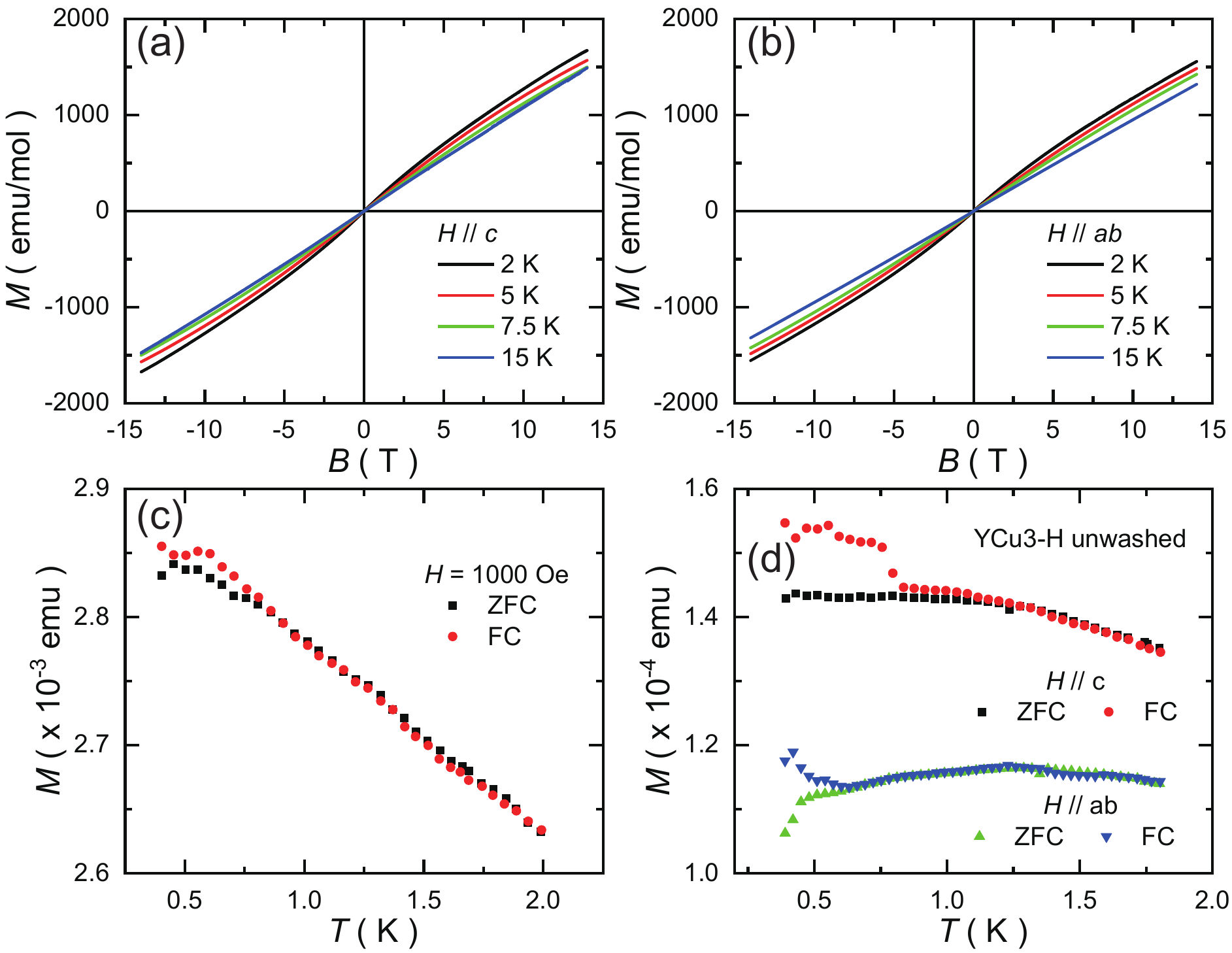}
 \caption{(a) and (b) The $M-H$ loops for YCu3-H at different temperatures for $H // c$ and $H // ab$, respectively. (c) The temperature dependence of $M$ at 1000 Oe. The samples were randomly oriented to put more samples in the capsule to gain better signal. (d) The temperature dependence of $M$ at 1000 Oe for the unwashed YCu3-H samples. The backgrounds in the measurements are different for different field directions.}
 \label{chi}
\end{figure}

Figure \ref{chi}(a) shows the $M-H$ loop for $H // c$, where no hysteresis is found at all temperatures. At 2 K, the slope $dM/dH$ decreases with increasing field and becomes almost field independent above about 5 T. At 15 K, $M$ linearly depends on $H$ for the whole field range. Similar behaviors are also observed for $H // ab$, as shown in Fig. \ref{chi}(b). The anisotropy $M_c/M_{ab}$ between $H // c$ and $H // ab$ at high fields is about 1.07 at 2 K and slightly increases with increasing temperature ( $\sim$ 1.13 at 15 K ). 

To further investigate the magnetic susceptibility at lower temperatures, we studied the temperature dependence of the magnetization below 2 K, as shown in Fig. \ref{chi}(c). No divergent behavior is found, suggesting the absence of free or weakly correlated spins. A slight difference between the zero-field-cooling (ZFC) and field-cooling (FC) processes appears below about 0.8 K. The ratio between this difference $\Delta M$ and the mean value of $M$ is less than 0.6\%, suggesting it may not come from the intrinsic properties of the samples. In fact, the measurements of the unwashed YCu3-H samples reveal similar behaviors with $\Delta M/M >$ 6\%, strongly demonstrating that the difference between ZFC and FC come from the external magnetic impurities, most of which were attached to the crystal surfaces. Overall, the intrinsic magnetic susceptibility of our samples should tend to a constant when $T$ goes to zero.

The above results demonstrate that the YCu$_3$(OH)$_6$Br$_2$[Br$_{x}$(OH)$_{1-x}$] system indeed has few magnetic impurities, as expected \cite{ChenXH20}. As shown in the Supplemental Material \cite{note1}, the magnetic impurities in our samples are mostly attached to the sample surfaces and can be removed by ultrasonic washing in water. For the washed samples, the existence of a very small amount of impurities has negligible effects on determining the thermodynamical properties as shown by both the specific-heat and magnetic-susceptibility results. This is different from herbertsmithite and many other KHAs \cite{VriesMA08,FreedmanDE10,KermarrecE11,LIY14,FengZL17,FengZL18b,YYHuang2021,YingFu2021}, for which the effect of magnetic impurities is very hard to separate from the bulk properties.

We thus conclude that  YCu$_3$(OH)$_6$Br$_2$[Br$_{x}$(OH)$_{1-x}$] is a strong candidate for realizing the Dirac QSL state, where the low-energy spinons form Dirac cones and their interactions are mediated by the emergent gauge fields~\cite{DHKim1997,RanY07,HermeleM08,XYXu2019}. The most promising evidence comes from the specific-heat results. First, the specific heat shows a $T^2$ dependence at zero field, which has been predicted for a U(1) Dirac QSL because of the Dirac nodes \cite{RanY07}. Second, the linear temperature dependence of the specific heat is found at high fields, suggesting the appearance of  the spinon Fermi surface, which has also been predicted theoretically \cite{RanY07}. It is interesting to note that this linear component shows up only above 2 T, which is again consistent with the condition for the above theoretical argument, i.e., $k_B T \ll \mu_B B$. Moreover, the linear-field-dependence and quadratic-temperature-dependence coefficients for YCu3-H are 0.01 J/mol TK$^2$ and 0.33 J/mol K$^3$, respectively, which gives a ratio that is just about 1/7 of the theoretical predicted value ($\sim$ 0.21 T$^{-1}$) 	\cite{RanY07}. For YCu3-D, the ratio is similar. These results may give a hint of the detailed structure of the Dirac nodes and low-energy spinon excitations and could even imply some kind of gauge fluctuation. It is particularly interesting to point out that the low-temperature specific heats behave differently for the YCu3-H and YCu3-D samples, which suggests that the fine structures of Dirac cones may be tuned. It is also worth noting that although the magnetic susceptibility $\chi$ is expected to show a linear temperature dependence, it tends to a large constant when $T$ goes to zero for our sample. Theoretically, it has been shown that a non-zero $\chi$ at 0 K is indeed possible for a gapless QSL \cite{ZhitomirskyME02,ZhouY08,SakaiaT18,ChenX18,BernuB20}.  Our observations therefore provide a promising experimental signature of the material realization of the highly non-trivial phase with emergent matter fields (Dirac spinon) coupled with gauge fields~\cite{DHKim1997,RanY07,HermeleM08,AssaadFF16,GazitS17,XYXu2019}, which has been pursued by broad communities ranging from quantum material to high energy.

It is interesting that a Dirac QSL may be realized in our sample but not in YCu$_3$(OH)$_6$Cl$_3$, which exhibits an AFM order at low temperatures \cite{ZorkoA19,ZorkoA19b,BarthelemyQ19}. This order is shown to be caused by the large Dzyaloshinskii–Moriya (DM) interaction, which also gives rise to a hump in $C/T$ at 16 K \cite{ArhT20}. In our case, the hump is at about 1.5 K, which means that the DM interaction is either an order smaller or even absent since the hump could also come from the low-energy excitations for a KHA \cite{SchnackJ18}. This is also consistent with the negligible magnetic anisotropy shown by the $M-H$ loops. The small DM interaction may be associated with the random distributions of Y and Br2 atoms. We note that the YCu$_3$(OH)$_6$Cl$_{3-x}$ ($x$ = 1/3) system has been suggested to show no magnetic order \cite{BarthelemyQ19}, which may be connected to our samples with partial occupancy of Br2 and two sites for Y, as shown in Tables I and II. However, no distortion for Cu is found for YCu$_3$(OH)$_6$Br$_2$[Br$_{x}$(OH)$_{1-x}$].

In conclusion, we have shown that the magnetic ground state of the YCu$_3$(OH)$_6$Br$_2$[Br$_{x}$(OH)$_{1-x}$] system may be a Dirac QSL on a 2D kagome lattice using the low-temperature specific-heat measurements. The very small number of magnetic impurities make it possible to directly compare theoretical and experimental results, helping us realize  KHA models in a real material. Moreover, the fine structure of Dirac nodes may be tuned by site disorders, which needs to be further studied.

Note added. Recently, we noted that studies on single-crystal YCu$_3$[OH(D)]$_{6.5}$Br$_{2.5}$ have also been reported \cite{LiuJ22}.

We thank Prof. T. Xiang, Y. Zhou and Y. Qi for helpful discussions. This work is supported by the National Key Research and Development Program of China (Grants No. 2017YFA0302903, No. 2021YFA1400401, No. 2016YFA0300502, No. 2020YFA0406003, No. 2017YFA0303100), the National Natural Science Foundation of China (Grants No. 11961160699, No. 11874401, and No. 11822411), the K. C. Wong Education Foundation (Grants No. GJTD-2020-01), the Strategic Priority Research Program (B) of the Chinese Academy of Sciences (Grants No. XDB33000000, and No. XDB25000000),  and the RGC of Hong Kong SAR of China (Grant No. 17303019, No. 17301420, No. 17301721, and No. AoE/P-701/20). H. L. is grateful for the support from the Youth Innovation Promotion Association of CAS (Grant No. Y202001) and the Beijing Natural Science Foundation (Grant No. JQ19002).

\end{document}